\documentclass[letterpaper,12pt]{article}   %% LaTeX 2e (preferred)
\usepackage{osajnl2} %% do not use with REVTeX4
\usepackage{hyperref} %% optional
\usepackage{amsmath,amsfonts,amssymb, wrapfig}
\usepackage[mediumspace,mediumqspace,Grey,squaren]{SIunits}

\newcommand{\diff}{\text{d}}

\begin{document}

\title{Inverse scattering of 2d photonic structures by
  layer-stripping}

%% For REVTeX it is possible to automate superscript and e-mail callouts with the superscriptaddress option; see REVTeX4 documentation.

\author{Marte P. Hatlo Andresen,$^{1}$ Harald E. Krogstad$^1$, and Johannes Skaar$^{2,3,*}$}
\address{$^1$Department of Mathematical Sciences, Norwegian University of
  Science and Technology, \\ NO-7491 Trondheim, Norway}
\address{$^2$Department of Electronics and Telecommunications, Norwegian University of Science and
Technology, \\ NO-7491 Trondheim, Norway}
\address{$^3$University Graduate Center, \\ NO-2027 Kjeller, Norway}
\address{$^*$Corresponding author: johannes.skaar@iet.ntnu.no}

\begin{abstract}

Design and reconstruction of 2d and 3d photonic structures are usually carried
out by forward simulations combined with optimization or intuition.
Reconstruction by means of layer-stripping has been applied in seismic
processing as well as in design and characterization of 1d photonic structures
such as fiber Bragg gratings. Layer-stripping is based on causality, where the
earliest scattered light is used to recover the structure layer-by-layer. 

Our set-up is a 2d layered nonmagnetic structure probed by plane polarized
harmonic waves entering normal to the layers. It is assumed that the
dielectric permittivity in each layer only varies orthogonal to the
polarization.  Based on obtained reflectance data covering a suitable
frequency interval, time-localized pulse data are synthesized and applied to
reconstruct the refractive index profile in the leftmost layer by identifying the
local, time-domain Fresnel reflection at each point. Once the first layer is
known, its impact on the reflectance data is stripped off, and the procedure
repeated for the next layer.  

Through numerical simulations it will be demonstrated that it is possible to
reconstruct structures consisting of several layers. The impact of evanescent
modes and limited bandwidth is discussed.

\end{abstract}

% REPLACE WITH CORRECT OCIS CODES FOR YOUR ARTICLE
                          % NOTE: \ocis{} IS ALIASED TO \pacs{} BUT MUST
                          % FORMAT THE TERMS CORRECTLY FOR EACH JOURNAL

\maketitle %% null function with osajnl.sty

\section{Introduction}

Photonic crystals have been an exciting field of research since Eli
Yablonovitch and Sajeev John published their papers 
 in 1987 \cite{yablonovitch_87, john}. Photonic crystals are
periodic structures designed to affect the propagation of
electromagnetic waves \cite{joannopoulos}. The usual way of
constructing such optical components is to carry out numerical
simulations of electromagnetic waves hitting and being scattered off a
trial design. The design is then changed by optimization or intuition until the
structure has the desired properties. In this paper we use a
layer-stripping procedure to show that it is possible to reconstruct a
photonic structure from a set of reflectance data based on harmonic waves. In other words, we will
look at how the inverse problem can be solved. The ultimate goal is to
be able to reconstruct the structures from the observations and use
this as a supplementary tool for the design of structures with desired features.

The idea of layer-stripping is to reconstruct the properties of a medium from scattered data originating from an emitted wave pulse on the
boundary of the domain. Layer-stripping is based on causality,
where the earliest scattered wave from each depth layer is used to
recover the structure layer-by-layer.  The method was first used in solving inverse
acoustic scattering problems for seismic data; overviews can be found
in \cite{bruckstein,yagle}. Later, the method has also been applied to
the design and characterization of one-dimensional photonic crystal
structures such as fiber Bragg gratings \cite{skaar2} and multimode
structures \cite{waagaard}. 

In the present simulation study, we are considering a two-dimensional structure, layered normal to 
the $z$-direction, with periodically varying refractive index in the
$x$-direction while being constant in the $y$-direction. The structure
is probed by plane polarized harmonic electromagnetic waves traveling in the
$z$-direction, with the electric field vector pointing in the
$y$-direction, thus representing a two-dimensional wave propagation problem. 

The forward problem, consisting of obtaining reflectance data for a known
structure, may be formulated in terms of transfer matrices, providing
reflection and transmission matrices for each layer as well as the whole
structure. 
The corresponding inverse problem, consisting of
recovering the properties of the unknown structure from the reflectance data,
requires data for several wave numbers and a range of frequencies.

The key step of the synthetic layer-stripping algorithm is to combine
reflectance data for different frequencies so as to synthesize data from a
short pulse at the time the pulse hits the surface. The data may then be
applied to reconstruct the refractive index profile in the leftmost layer by
identifying the local, time-domain Fresnel reflection at each point. Once the
first layer is known, its transfer matrix may be
computed, and the impact of the layer on the reflectance data eliminated. The
same idea is then applied to the modified reflectance data, and continuing in
the same way we are, at least in principle, able to reconstruct the whole
structure.

\section{The forward problem}
\label{sec:forward}

We are considering photonic crystals consisting of homogeneous, dielectric and
nonmagnetic ($\mu=\mu_{0}$) materials, where the dielectric permittivity,
$\epsilon\left(  x,z\right)  $, is varying in the $x$- and
$z$-directions, and being constant in the $y$-direction. The structure
consists of layers orthogonal to the $z$-axis, with $\epsilon$ 
constant with respect to $z$ (and $y$) in each
layer. We assume that the structure contains $N$ \ layers of finite thickness
occupying the space from $z_{0}=0$ to $z_{N}$. Layer $i$ spans $z_{i-1}\leq
z\leq z_{i}$, with a thickness $\Delta_{i}=z_{i}-z_{i-1}$, and permittivity
$\epsilon_{i}(x)$. Outside the structure we assume that it is vacuum. By inserting the Fourier transform in time of the
electric and magnetic fields, e.g. 
\begin{equation}
\mathbf{E}\left(  \mathbf{x,}t\right)  =\frac{1}{2\pi}\int_{-\infty}^{\infty
}\mathbf{e}\left(  x,z,\omega\right)  e^{-i\omega t}\mathrm{d}\omega,
\end{equation}
and similarly for $\mathbf{H}$, into Maxwell's equations, we obtain
\begin{eqnarray}
\nabla\times\mathbf{e}(x,z,\omega)-i\omega\mu_{0}\mathbf{h}
(x,z,\omega)  &=&0,\label{MX5}\\
\nabla\times\mathbf{h}(x,z,\omega)+i\omega\epsilon_{0}
\epsilon(x,z)\mathbf{e}(x,z,\omega)  &=&0,\label{MX6}\\
\nabla\cdot(\epsilon(x,z)\mathbf{e}(x,z,\omega))
&=&0,\label{MX7}\\ 
\nabla\cdot\mathbf{h}(x,z,\omega)  &=&0. \label{MX8}
\end{eqnarray}
In addition, from the standard continuity conditions for $\mathbf{E}$
and $\mathbf{H}$,  parallel ($\Vert$) and orthogonal ($\bot$) to the layer
boundaries, the following conditions have to apply
\begin{eqnarray}
\epsilon_{1}\mathbf{e}_{1}^{\perp}  &=& \epsilon_{2}\mathbf{e}
_{2}^{\perp},\\
\mathbf{e}_{1}^{\parallel}  &=& \mathbf{e}_{2}^{\parallel},\\
\mathbf{h}_{1}  &=& \mathbf{h}_{2}.
\end{eqnarray}
Omitting the dependence of $\omega$, direct substitution shows that
the Fourier transformed Maxwell's 
equations in the present case admit solutions of the form
\begin{eqnarray}
\mathbf{e}(x,z)  &=& e\left(  x,z\right)  \mathbf{\hat{y}
,}\label{p1}\\
\mathbf{h}(x,z)  &=& \frac{1}{i\omega\mu_{0}}\left(
-\frac{\partial e\left(  x,z\right)  }{\partial z}\mathbf{\hat{x}+}
\frac{\partial e\left(  x,z\right)  }{\partial x}\mathbf{\hat{z}}\right)
\mathbf{,} \label{p2}
\end{eqnarray}
provided \smallskip$e\left(  x,z\right)  $ satisfies the scalar Helmholtz
equation,
\begin{equation}
\frac{\partial^{2}e}{\partial x^{2}}+\frac{\partial^{2}e}{\partial z^{2}
}+\epsilon(x,z)k^{2}e=0, \label{starteq}
\end{equation}
where $k=\omega/c_{0}$, $c_{0}^{2}=1/\mu_{0}\epsilon_{0}$, and $e$ as as
well as its partial derivatives are continuous across the layer boundaries.

The structure will be probed using the plane polarized
waves in Eqs. (\ref{p1}) and (\ref{p2}), and all transmitted and
reflected waves will have the same polarization. 

The direct problem consists of solving Eq. (\ref{starteq}) with appropriate
boundary conditions. These will be plane waves entering from the half space
$z\leq0$ (thus, having positive $z$-component wave numbers), whereas no waves
are supposed to enter the structure from the region $z>z_{N}$.

\section{Solution by transfer matrices} 
\label{sec:transfer}

Let $L$ be the extension of the computational domain in the $x$-direction. By imposing $L$-periodic boundary conditions, we may
expand
the electric field and the permittivity  
into the Fourier series
\begin{eqnarray}
\label{h-field}
\mathbf{e}(x,z) = e(x,z)\mathbf{\hat{y}} &=& \sum_m E^{(m)}(z) \exp(i k_{x}^{(m)}x)
\mathbf{\hat{y}}, \\
\epsilon(x,z) &=& \sum_m \epsilon^{(m)}(z) \exp(i k_{x}^{(m)}x),
\end{eqnarray}
where $k_{x}^{(m)} = 2 \pi m/L$. The Fourier components are given as 
\begin{equation}
\label{eps_m}
\epsilon^{(m)}(z) = \frac{1}{L} \int_0^L \epsilon(x,z) \exp(-i k_{x}^{(m)}
x) \mathrm{d}x.
\end{equation}
By inserting the Fourier series into Helmholtz' equation we obtain
\begin{equation}
\label{helm}
\frac{\mathrm{d}^2 E^{(m)} (z)}{\mathrm{d} z^2} - \left(k_{x}^{(m)}\right)^2
E^{(m)}(z) + k^2  \sum_{m'} \epsilon^{(m-m')} (z)E^{(m')}(z)=0,
\end{equation}
for each $m$. Let $\mathbf{E} = \{E^{(m)}\}_{m\in \mathbb{Z}}$  and use
$k^2 = \left(k_{x}^{(m)}\right)^2 + \left(k_{z}^{(m)}\right)^2$. We may then write Eq. (\ref{helm}) as the
matrix equation 
\begin{equation}
\label{eq5}
\frac{\mathrm{d}^2 \mathbf{E} (z)}{\mathrm{d} z^2} + \left(\mathbf{k}_z^2  +
  \mathbf{V}(z) \right) \mathbf{E}(z) = 0,
\end{equation}
where $\mathbf{k}_z = \mathrm{diag}(k_{z}^{(m)})$, and $\mathbf{V}(z)$ is the
infinite dimensional Toeplitz matrix operator defined as
\begin{equation}
\label{Toep_op}
\mathbf{V}(z) = -k^2 \mathbf{I} + k^2  
\left[
\begin{array}[c]{ccccc}
\ddots & \vdots & & & \\
\cdots & \epsilon^{(0)} & \epsilon^{(-1)} & \epsilon^{(-2)} & \\
 & \epsilon^{(1)} &  \epsilon^{(0)} & \epsilon^{(-1)} & \\
 & \epsilon^{(2)} &  \epsilon^{(1)} &  \epsilon^{(0)} & \cdots\\
 & & & \vdots & \ddots
\end{array}
\right].
\end{equation}
Eq. (\ref{eq5}) may be decomposed into a first order system by writing
$\mathbf{E}(z) = \mathbf{E^+}(z) + \mathbf{E^-}(z)$, and require %. In fact, the equations
\begin{subequations}\label{decomposition}
\begin{eqnarray}
  \frac{\diff \mathbf{E^+}(z)}{\diff z} &=& i\mathbf{k}_z\mathbf{E^+}(z) +
  i (2 \mathbf{k}_z)^{-1} \mathbf{V}(z)
  (\mathbf{E^+}(z)+\mathbf{E^-}(z)),\label{decomp1}\\ 
  \frac{\diff \mathbf{E^-}(z)}{\diff z} &=& -i\mathbf{k}_z\mathbf{E^-}(z) -
  i (2\mathbf{k}_z)^{-1} \mathbf{V}(z)(\mathbf{E^+}(z)+\mathbf{E^-}(z)).\label{decomp2}
\end{eqnarray}
\end{subequations}
Eq. (\ref{eq5}) follows easily from Eqs. (\ref{decomp1}) and (\ref{decomp2})   
after summation and differentiation. 
Note that outside the structure $\mathbf{V}$ equals zero, and the
solution to Eq. (\ref{decomposition}) are forward and backward going
waves. 

By writing
\begin{equation}
\mathbf{\Psi}(z)   \mathbf{=}\left[
\begin{array}
[c]{c}%
\mathbf{E^+}(z)\\
\mathbf{E^-}(z)
\end{array}
\right]  , \quad
\mathbf{C}(z)   \mathbf{=}\left[
\begin{array}
[c]{cc}%
i \mathbf{k}_z + i (2 \mathbf{k}_z)^{-1} \mathbf{V}(z) & i (2
\mathbf{k}_z)^{-1} \mathbf{V}(z)\\
 - i (2 \mathbf{k}_z)^{-1} \mathbf{V}(z) &-i \mathbf{k}_z- i (2
\mathbf{k}_z)^{-1} \mathbf{V}(z)\\
\end{array}
\right]  , \label{c}
\end{equation}
Eq. (\ref{decomposition}) can be brought into the compact matrix form
\begin{equation}
\label{diff_mat}
\frac{\diff \mathbf{\Psi}(z)}{\diff z} = \mathbf{C}(z) \mathbf{\Psi}(z).
\end{equation}
Since the permittivity is independent of $z$ within a
layer, the matrix $\mathbf{C}$ will be constant for
each layer.  Thus, Eq. (\ref{diff_mat}) can be integrated to 
\begin{equation}
\mathbf{\Psi}(z_b) = \exp[(z_b-z_a)\mathbf{C}]\mathbf{\Psi}(z_a)
\end{equation}
for $z_a$ and $z_b$ inside the same layer. Let $\mathbf{C}_i$ be the
$\mathbf{C}$-matrix for layer $i$. By defining
\begin{equation}
\label{Mi}
\mathbf{M}_i = \exp (\Delta_i \mathbf{C}_i ),
\end{equation}
we obtain
\begin{equation}
\label{psi}
\mathbf{\Psi}(z_i) = \mathbf{M}_i\mathbf{\Psi}(z_{i-1}).
\end{equation}
Since $e$ and $\partial e/\partial
z$ are continuous across the layer boundaries, $\mathbf{E^+}$ and
$\mathbf{E^-}$ also must be continuous here. By successive
applications of Eq. (\ref{psi}) we are able to propagate through the
structure from $z_0$ to $z_N$,  
\begin{equation}
\mathbf{\Psi}(z_N) = \prod_{i=N}^0 \mathbf{M}_i \mathbf{\Psi}(z_0) =
\mathbf{M} \mathbf{\Psi}(z_0).
\end{equation}
The boundary conditions at $z=z_0=0$ are contained in
\begin{equation}
\mathbf{\Psi}(z_0)=\left[
\begin{array}
[c]{c}%
\mathbf{E^+}(z_0)\\
\mathbf{E^-}(z_0)
\end{array}
\right]  ,
\end{equation}
where $\mathbf{E^+}(z_0)$ is given by the probing waves, and
$\mathbf{E^-}(z_0)$ is unknown. At the other end of the structure,
\begin{equation}
\mathbf{\Psi}(z_{N})=\left[
\begin{array}
[c]{c}
\mathbf{E^+}(z_N)\\
0
\end{array}
\right]  ,
\end{equation}
where $\mathbf{E^+}(z_N)$ has to be determined. If we partition the
matrix $\mathbf{M} $ according to the definition of $\mathbf{\Psi}$, 
\begin{equation}
\mathbf{M} =\left[
\begin{array}
[c]{cc}
\mathbf{M}_{11} & \mathbf{M}_{12}\\
\mathbf{M}_{21} & \mathbf{M}_{22}
\end{array}
\right]  ,
\end{equation}
and consider the matrix equation
\begin{equation}
\left[
\begin{array}
[c]{c}
\mathbf{E^+}(z_N)\\
0
\end{array}
\right]  =\left[
\begin{array}
[c]{cc}%
\mathbf{M}_{11} & \mathbf{M}_{12}\\
\mathbf{M}_{21} & \mathbf{M}_{22}
\end{array}
\right]  \left[
\begin{array}
[c]{c}%
\mathbf{E^+}(z_0)\\
\mathbf{E^-}(z_0)
\end{array}
\right]  , \label{L0}
\end{equation}
the formal solution is easily seen to be
\begin{eqnarray}
\mathbf{E^-}(z_0)  &=& \mathbf{R}\mathbf{E^+}(z_0),\label{R}\\
\mathbf{E^+}(z_N)  &=& \mathbf{T}\mathbf{E^+}(z_0), \label{T}
\end{eqnarray}
where $\mathbf{R}$ and $\mathbf{T}$ are the so-called \emph{reflection} and
\emph{transmission matrices} for the structure,
\begin{eqnarray}
\mathbf{R}  &=& -\mathbf{M}_{22}^{-1}\mathbf{M}_{21},\\
\mathbf{T}  &=& \mathbf{M}_{11}-\mathbf{M}_{12}\mathbf{M}_{22}^{-1}%
\mathbf{M}_{21}.
\end{eqnarray}
The reflection and transmission matrices provide $\mathbf{E^-}(z_0)$
and $\mathbf{E^+}(z_N)$ for all possible input $\mathbf{E^+}(z_0)$. 

To solve the inverse problem, we need reflectance data for all
possible  incident wave numbers. Such a set of excitation-response
pairs can be described by the following equation  
\begin{equation}
\left[
\begin{array}
[c]{c}%
\mathbf{T}\\
\mathbf{0}
\end{array}
\right]  = 
\mathbf{M} 
\left[
\begin{array}
[c]{c}%
\mathbf{I}\\
\mathbf{R}
\end{array}
\right].
\label{RT_eq}
\end{equation}
Here, each column $i$ of $\mathbf{I}$ corresponds to an experiment where
the incident field amplitude is $1$ for one of the Fourier components
and zero for the others. The $i$th column of $\mathbf{R}$ is
the reflection at $z_0$, and the $i$th column of $\mathbf{T}$
is the corresponding transmission at $z_N$.

\section{The inverse problem}
\label{sec:inv_prob}

The inverse problem is solved by combining reflectance data for a range of
frequencies, $\left[  \omega_{1},\omega_{2}\right]  $, so as to synthesize data
from a short pulse at the time the pulse hits the surface. The permittivity of
the layer may then be recovered as described below if the pulse width in
time, $\mathcal{O}\left( 10 \left(  \omega_{2}-\omega_{1}\right)  ^{-1}\right)
$, is shorter than the round-trip travel time in the layer, $2\Delta_{i}%
/c$. In this study we shall, for simplicity, assume that the layer thicknesses,
$\Delta_{i}$, are known. It is, in principle, possible to do the layer-stripping without knowledge of the layer thickness. This will be briefly discussed this in Sec. \ref{num_ex}. From the permittivity and the thickness it is now
possible to compute the transfer matrix for the layer,
and the impact of the layer removed from the reflection data. Thus, we obtain
reflectance data for the same structure, but without the leftmost layer, and may
then repeat the steps above until the entire structure has been reconstructed.

\subsection{Computing the permittivity}

\begin{figure}[t]
  \centerline{\includegraphics[width=0.75\textwidth]{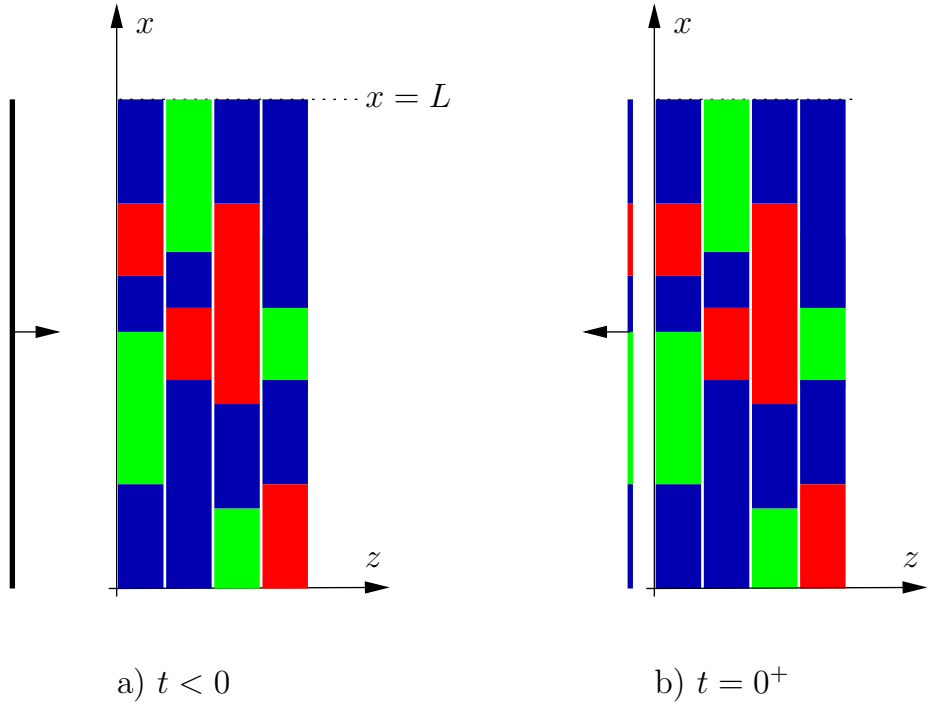}}
  \caption{(a) A plane wave pulse is incident to a layered 2d structure. For the structure different colors indicate different refractive indices. (b) Immediately after $t=0$ the field has only been affected by the first layer; thus we may identify the first layer from the first part of the reflected field in the time-domain. AKSFig1.eps.}
\label{f1:1lp}
\end{figure}

Let us assume that we have reflectance data from an
experiment where the incident wave is a plane wave pulse,
$F(t-z/c_0)$, see Fig. \ref{f1:1lp}. The incident wave enters from $z<0$, and first hits the
leftmost layer of the structure. The wave speed in the first layer is $c_1(x)$.

Let us consider a small neighborhood around the point $x$, and
assume that the permittivity in the leftmost layer is varying slowly enough,
so that we may take it to be constant and equal to $\epsilon_1(x)$
inside the area we consider. Furthermore, we restrict our calculation
to a small time interval so that the reflected and transmitted waves
are only affected by the first layer.
 
At the left side of the boundary, the electric field will be a sum of the incoming
and the reflected wave, and at the right side, the field consists of
the transmitted wave. Thus, continuity of the transversal
electromagnetic fields gives us  
\begin{equation}
\label{rt_bon}
F(t-\frac{z}{c_0}) + R_1(x) F(t+ \frac{z}{c_0}) = T_1(x)
F(t-\frac{z}{c_1(x)}), 
\end{equation}
and
\begin{equation}
\label{rt_bon_d}
-\frac{1}{c_0} F'(t-\frac{z}{c_0}) + \frac{R_1(x)}{c_0}
F'(t+\frac{z}{c_0}) = -\frac{T_1(x)}{c_1(x)}F'(t-\frac{z}{c_1(x)}), 
\end{equation}
for some reflection coefficient $R_1(x)$. Obviously, the incident wave pulse $F$ needs to be
short for Eqs. (\ref{rt_bon}) and (\ref{rt_bon_d}) to hold. 
For $z=0$ and $0 \leq t \ll \Delta_1/c_0$ we thus have from
Eqs. (\ref{rt_bon}) and (\ref{rt_bon_d}) 
\begin{eqnarray}
1+R_1(x) &=& T_1(x),  \\
-1+R_1(x) &=& -T_1(x)\frac{c_0}{c_1(x)}, 
\end{eqnarray}
leading to the following expression for the permittivity in the leftmost layer:
\begin{equation}
\label{eps_uttrykk}
\epsilon_1(x) = \epsilon_0  \left( \frac{1-R_1(x)}{1+R_1(x)} \right)^2. 
\end{equation}
Eq. \eqref{eps_uttrykk} can be interpreted as a local Fresnel equation; it connects the local reflection coefficient $R_1(x)$ and the local permittivity $\epsilon_1(x)$. 

\subsection{Layer-stripping}
\label{layerstrippth}

For the layer-stripping we shall assume that the layer thickness is known.  
Once we have computed $\epsilon_1(x)$, we obtain the transfer matrix $\mathbf{M}_{1}$ from Eq. (\ref{Mi}).
As in Eq. (\ref{psi}), the forward and backward going waves just before
the second layer can be computed as 
\begin{equation}
\left[
\begin{array}
[c]{c}
\mathbf{E^+}(z_1)\\
\mathbf{E^-}(z_1)
\end{array}
\right]  =\mathbf{M}_1\left[
\begin{array}
[c]{c}
\mathbf{E^+}(z_0)\\
\mathbf{E^-}(z_0)
\end{array}
\right] =\mathbf{M}_1\left[
\begin{array}
[c]{c}
\mathbf{I}\\
\mathbf{R}
\end{array}
\right]  .\label{1L}
\end{equation}
Note that here $\mathbf{E^+}$ and $\mathbf{E^-}$ are matrices as in
Eq. (\ref{RT_eq}). Similar to the solution for the full structure,
Eqs. (\ref{R}) and (\ref{T}), we may define 
\begin{eqnarray}
\mathbf{E^+}(z_N) &=& \mathbf{\tilde{T}}\mathbf{E^+}(z_1),\\
\mathbf{E^-}(z_1) &=& \mathbf{\tilde{R}}\mathbf{E^+}(z_1), \label{R1}
\end{eqnarray}
and thus the new reflection matrix can be found as
\begin{equation}
\label{R_0}
\mathbf{\tilde{R}} = \mathbf{E^-}(z_1) \mathbf{E^+}(z_1)^{-1}.
\end{equation}

\section{Algorithm}
\label{sec_alg}

For the numerical experiments, which we will discuss in Sec. \ref{num_ex}, 
we first solve the forward problem to obtain valid reflectance data $\mathbf{R}(k_{x}^{(m')},k_{x}^{(m)}, \omega)$ for each of $N_{\omega}$ frequencies in the span $[\omega_1,\omega_2]$. Here $k_{x}^{(m')}$ denote the incident wave numbers, and $k_{x}^{(m)}$ the reflected wave numbers. 

For the inverse problem, the first step is to synthesize a time localized pulse in space and time to be applied in Eq. (\ref{eps_uttrykk}).
Only data for incident waves normal to the structure are used in
the identification of the layers.  We start with a transformation
from $k_{x}^{(m)}$ to $x$,  
\begin{equation}
\label{rxw}
r(x,\omega) = \sum_{m} \mathbf{R}(k_{x}^{(m')}=0,k_{x}^{(m)}, \omega) e^{i k_{x}^{(m)} x},
\end{equation}
where
\begin{equation}
k_{x}^{(m)} = \frac{2 \pi m}{L}, \quad m \in \mathbb{Z},\quad \text{and} \quad
\omega \in [\omega_1, \omega_2].
\end{equation}
Ideally, one would now carry out an inverse Fourier transform in order
to obtain the reflected field in the space- and time-domains resulting from an incident delta-pulse, 
\begin{equation}
\label{rxt}
R(x,t)=\frac{1}{2\pi}\int_{-\infty}^{\infty}r(x,\omega)e^{-i\omega
  t}\mathrm{d}\omega. 
\end{equation}
However, since we only know $r(x,\omega)$ in the frequency interval $\left[
\omega_{1},\omega_{2}\right]  $, 
we need to choose a window function $W(\omega)$ with support in that interval. The corresponding pulse in time is denoted $w(t)$, and the experience with some standard window functions are discussed in the next section.
 The synthetic response will then be  
\begin{equation}
\label{rwxt}
R_{w}(x,t)=
\frac{1}{2\pi}%
\int_{\omega_1}^{\omega_2}r(x,\omega)W(\omega)e^{-i\omega t}\mathrm{d}\omega.
\end{equation}
For $z=0$ and $0\leq
t\ll\Delta_{1}$, we see from Eq. (\ref{rt_bon}) that 
$R_{w}\left(  x,t\right)  =R_1\left(  x\right)  w\left(  t\right)
$. Hence, for $t=0$, we may write%
\begin{equation}
\label{r0x}
R_1\left(  x\right)  =\frac{R_w\left(  x,0\right)  }{w\left(  0\right)  }=
\frac{\int_{\omega_1}^{\omega_2} r(x,\omega
)W(\omega)\mathrm{d}\omega}{\int W(\omega)\mathrm{d}\omega}.
\end{equation}
The permittivity $\epsilon\left(  x\right)  $
is then obtained from Eq. (\ref{eps_uttrykk}). The layer-stripping has been discussed in Sec. \ref{layerstrippth}.

The numerical method may now be summarized in a pseudo code, as
follows.

\begin{itemize}
\item[] Input to the algorithm: Reflection matrices $\mathbf{R}(k_{x}^{(m')}=0,k_{x}^{(m)},\omega)$, for $\omega \in [\omega_1,\omega_2]$.
\item[1] Transform reflection matrices to the space/frequency domain using Eq. (\ref{rxw}). Next, determine the reflection coefficient at $t=0$ using Eqs.  (\ref{rwxt}) - (\ref{r0x}). The permittivity now follows from
  Eq. (\ref{eps_uttrykk}). 
\item[2] Use $\epsilon(x)$ and Eq. (\ref{eps_m}) to find the Fourier
  components. Then the transfer matrix can be computed from
  Eqs. (\ref{Toep_op}), (\ref{c}) and (\ref{Mi}).
\item[3] Knowing the transfer matrix, the forward and backward going waves before the next layer 
  can be computed from Eq. (\ref{1L}).
\item[4] Compute the new reflection matrix, using Eq. (\ref{R_0}).
\item[5] Return to step 1, to compute the permittivity in the next layer, until all layers have been found.
\end{itemize}

\section{Numerical experiments}
\label{num_ex}

For simplicity, we scale $\epsilon_0$ and $\mu_0$ (in free space)
to 1, and hence $c_0=1$. Now, the dispersion relation in free space
reduces to $k= \omega$, and the wavelengths will be $\lambda = 2
\pi / \omega$. The length of the synthesized pulse will depend on the frequency interval $[\omega_1, \omega_2]$, and be  of order $  2\pi\left(  \omega_{2}-\omega_{1}\right)  ^{-1}$. One must require that the pulse length is shorter than the  round-trip travel time in the first layer, i.e.
\begin{equation}
\frac{2\pi}{\omega_2 -\omega_1} \lessapprox \frac{2 \Delta_1}{c_1}.
\end{equation}
Thus, to be able to reconstruct the structure, we need to make sure that the frequency band is wide enough. Through simulations we have seen that for smaller frequency bands, the error fluctuates as the bandwidth grows. However, as soon as the frequency span is wide enough, the calculations are stable. 

As mentioned in Sec. \ref{sec_alg}, we use Eq. (\ref{RT_eq}) to compute reflection matrices. 
To compute the reflection matrix and solve the inverse problem, the number of Fourier components, $M$, the computational
domain, $L$, and the $N_{\omega}$ frequencies in the interval 
$[\omega_1, \omega_2]$, must be chosen. Now, the reflection and transmission matrices will have dimension $M \times M$ and the propagation matrix $\mathbf{M}$ dimension $2M \times 2M$. 
Note that given the resolution in the $x$-direction
\begin{equation}
\Delta x = \frac{L}{M},
\end{equation}
the maximum  Nyquist wave number  is given as
\begin{equation}
\max k_x = \frac{\pi M}{L}.
\end{equation}
Since
\begin{equation}
k_z = \sqrt{k^2-k_x^2} = \sqrt{\omega^2-k_x^2},
\end{equation}
evanescent modes in vacuum occur when $\max k_x > \omega$.

As we see in Eq. (\ref{R_0}), the updated reflection matrix is a product
of the backward traveling waves and the inverse of the forward traveling
waves. In a homogeneous medium, the forward and backward traveling
waves have the $z$-dependence $\exp(ik_z z)$ and $\exp (-ik_z z)$,
respectively. Thus, when we have evanescent modes with $k_z = iK$ for
some $K > 0$, we expect a noise amplification factor of the order of
$\exp(2K\Delta)$ when one layer is stripped off. Because of this, we 
restrict our frequency band to avoid a too large $K$.

Note that if the variations in the $x$-direction are fast, we will
need a finer resolution in $x$. 
To obtain this, we must either make $M$ larger, or $L$
smaller, which again implies that the maximum wave number, $\max k_x$,
increases. The result is that we need higher frequencies to avoid
evanescent modes. Thus, it is convenient to restrict the attention to
structures that 
vary slowly with respect to $x$. We will, however, in Example 2 see
that we are also able to reconstruct a structure with fast
variations. 

There are three effects that mainly contribute to the limitations in this
method. The first one is the effect of evanescent modes, which was 
discussed above. The second is related to the contrast between the
minimum and maximum refractive indices, $n_1$ and $n_2$. As the
contrast grow higher, more of the light is reflected by the first layers. This
will in turn make the calculations less accurate
\cite{bruckstein2,skaar3}. The last effect is also related to the bandwidth. As discussed in the beginning of this section, we need the pulse length to be smaller than the layer round-trip travel time, thus imposing a lower limit to the bandwidth. Conversely, if the bandwidth is fixed, the layer thicknesses have to be large enough.  

As discussed in Sec. \ref{sec_alg}, it is necessary to shape  the reflection data using some window function. Numerical experiments applying three different window functions on the interval
$\left[  \omega_{1},\omega_{2}\right]  $ are presented in
Sec. \ref{sec:ex1}. As expected, the 
experiments favor smoother windows over the simple rectangular window.
Consequently, a Hanning window has been applied for the rest of the
computations.

In all simulations we have assumed that we know the layer
thickness. It would, at least in principle, be possible to 
reconstruct the layer thickness. To do this, one need to use a
fictitious layer thickness, $\Delta z$, which must be small compared to
the expected layer thickness. Then this $\Delta z$ can be used in the
computations, to reconstruct each layer piece by piece. %, by
                
All the calculations where done using MATLAB on an Intel Core 2 Quad 2.83 GHz computer. For a typical calculation, the runtime was approximately 260 s. About 180 s was used on calculating the forward problem, i.e. the reflection matrices $\mathbf{R}(k_{x}^{(m')},k_{x}^{(m)},\omega)$, the calculation of the permittivity is neglectable, and the calculation and removal of the leftmost layer took about 80 s.                                

\subsection{Example 1}
\label{sec:ex1}

In this example we have chosen a structure where the permittivity within the layers is
given as
\begin{equation}
\epsilon(x) = \eta \pm \gamma \cos (\frac{x}{2}),
\end{equation}
and where the two different layers are obtained by
alternating between $+$ and $-$. The layer thickness has been set to 
$\pi/2$, and the
 parameters $\eta$ and $\gamma$ are chosen so that 
\begin{eqnarray}
\min \epsilon &=& n_1^2=1.0, \\
\max \epsilon &=& n_2^2,
\end{eqnarray}
i.e. $n_1$ is kept constant, while  $n_2^2$ is varied. 
For all the experiments presented here, $M=300$ and
$L=100$, giving a resolution in $x$ of $1/3$ and a maximum $k_x$ of
$3\pi$. The choice of frequency band depends on whether we want to include
evanescent modes in the computations,  how large $K$ we can tolerate, and the layer thickness. 

\begin{figure}[t]
  \centerline{\includegraphics[width=0.75\textwidth]{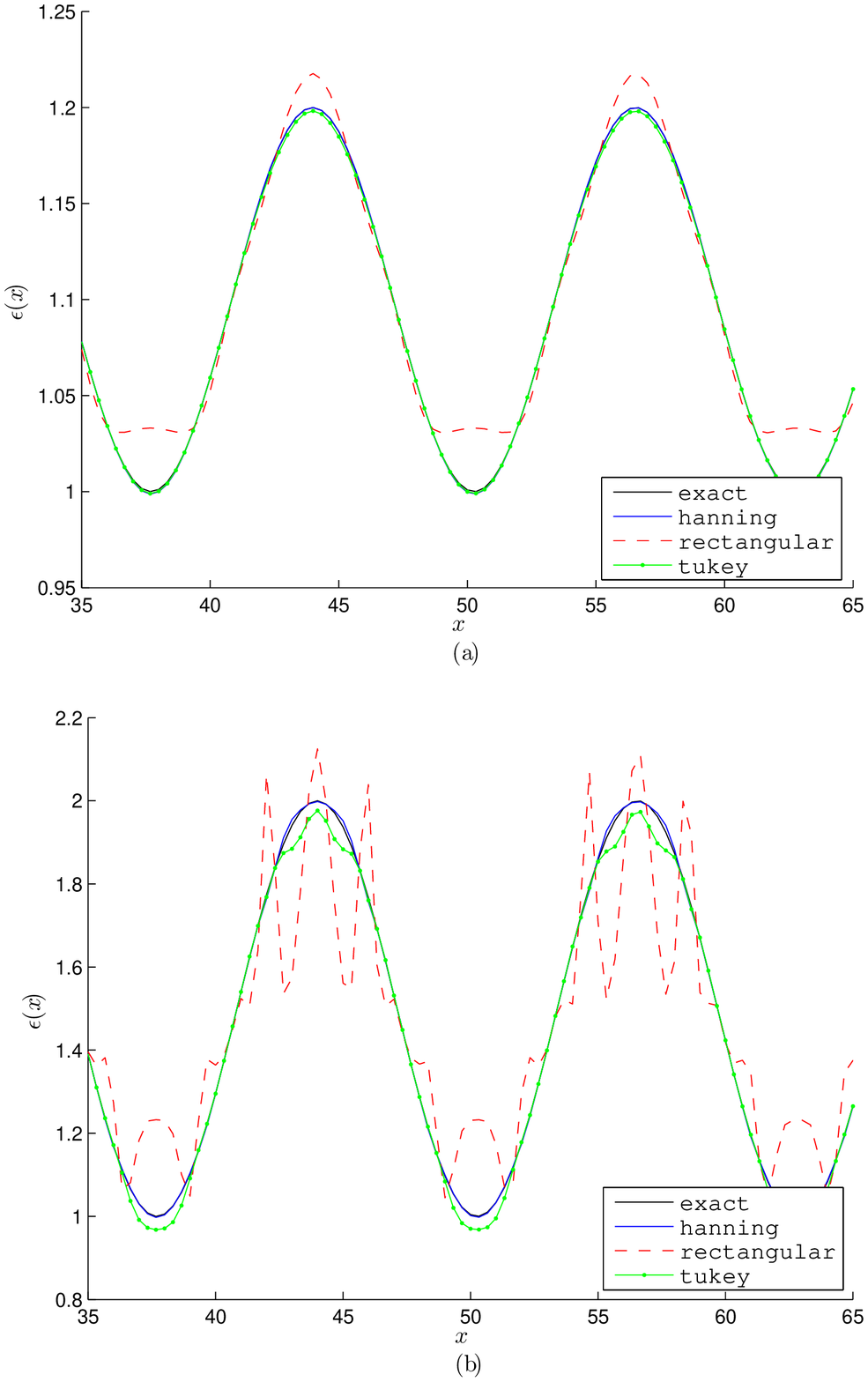}}
  \caption{Reconstruction of the second layer for different window
  functions, in (a)  $n_2^2=1.2$ and in (b)  $n_2^2=2.0$. Other parameters:
  $M=300$, $L=100$, 
$N_{\omega} =100$, $\omega \in[9,19]$, $\Delta z = \pi/2$, $n_1^2 =
1.0$. Note that only parts of the computational domain $L$ are shown in the plots. AKSFig2.eps.}
\label{f1:1}
\end{figure}

We first consider the choice of frequency window functions, Eq. (\ref{rwxt}), by testing 
three
different functions, a simple rectangular window, the Hanning window
 and the Tukey (tapered cosine) window. The
comparisons were carried out for two different values of $n_2$, 
$n_2^2 = 1.2$ for the first test, and $n_2^2 = 2.0$ for the second. The
real difference in the choice of window functions can only be seen in the
reconstruction of the second layer. Therefore, the results, displayed in Fig.  \ref{f1:1}, are only shown for this layer. The
rectangular window give poor results already for  
the lowest refractive index, whereas the two other choices give
a nice reconstruction. For the higher index, we see that also the
 Tucky window is beginning to give less
accurate results, while for the Hanning window there is almost no
visible difference from the exact curve. For the rest of the numerical
examples, we have therefore chosen to use the Hanning window.

\begin{figure}[t]
  \centerline{\includegraphics[width=0.75\textwidth]{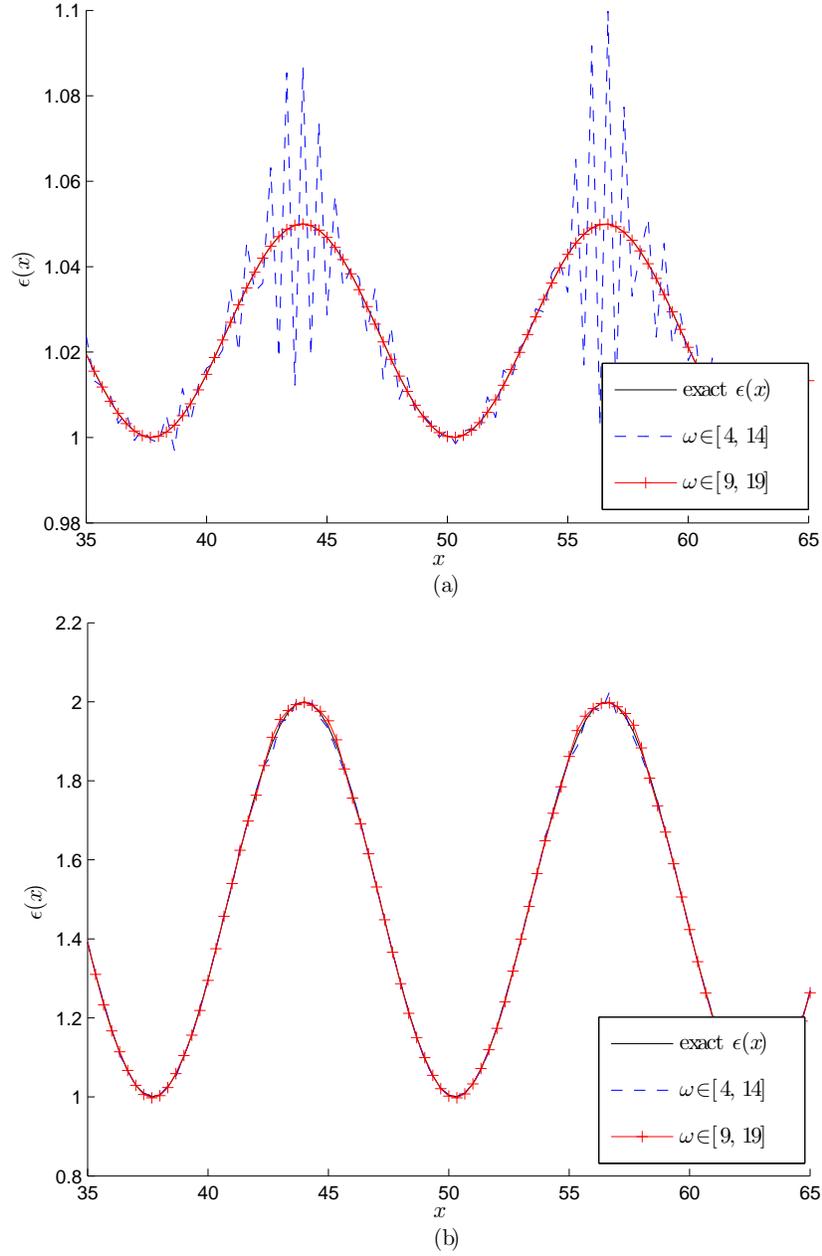}}
  \caption{Reconstruction of the second layer for different $\omega_1$
  and $\omega_2$, but with constant bandwidth. In (a) $n_2^2=1.05$,
  and in (b) $n_2^2=2.0$.  Other parameters:
  $M=300$, $L=100$, 
$N_{\omega} =100$, $\Delta z = \pi/2$, $n_1^2 = 1.0$. AKSFig3.eps.}
\label{f1:2}
\end{figure}

Let us now turn to  the effect of including evanescent
modes. In these experiments, the permittivity was first chosen such that
$n_1^2=1.0$ and $n_2^2 = 1.05$, and secondly we had $n_1^2=1.0$ and
$n_2^2=2.0$. The frequency bandwidth was set constant to $10$, but the lower
and upper frequencies were changed to include none, or some evanescent
modes.  Again, the first layer is nicely reconstructed, so we only
show results for the second layer in
Fig. \ref{f1:2}. Including some evanescent 
modes improves the results, but if $K$ gets too large, there are fluctuations
in the reconstruction. Note that as  $n_2$
gets higher, a larger $K$ is acceptable . This can be explained by the
fact that the field now is locally non-evanescent, since a local $k_z$
in the medium would be $k_z = (n^2 
\omega^2 -k_x^2)^{1/2}$.

\begin{figure}[t]
  \centerline{\includegraphics[width=0.75\textwidth]{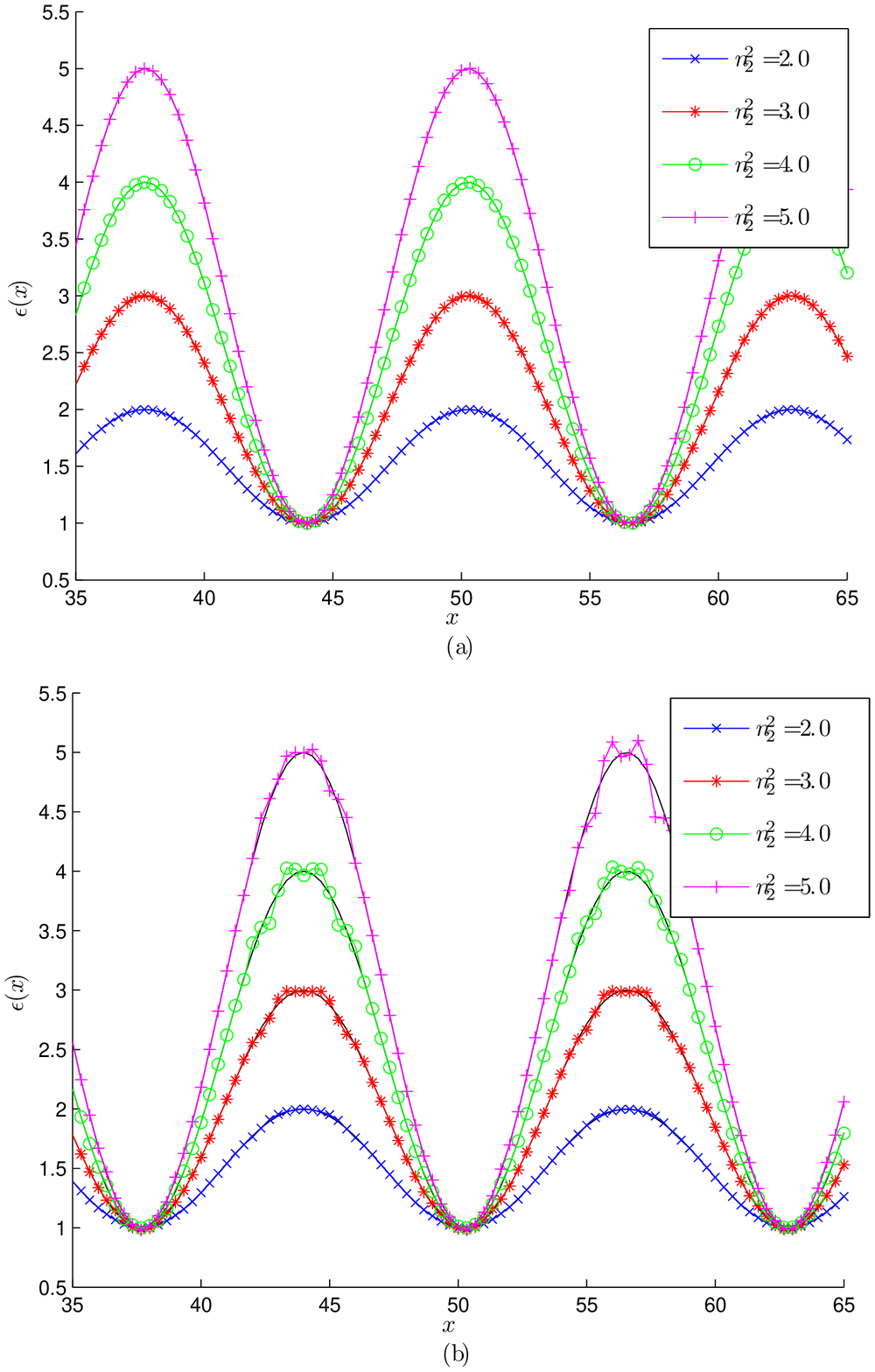}}
  \caption{Reconstruction of the first layer, (a), and the second
    layer, (b), for different choices of
  $n_2$.  Other parameters: $M=300$, $L=100$, $\omega \in [9,19]$, 
$N_{\omega} =100$, $\Delta z = \pi/2$, $n_1^2 = 1.0$. The solid black
lines represent the exact permittivity. AKSFig4.eps.}
\label{f1:3}
\end{figure}

Bigger contrasts in the refractive index may be simulated by keeping the lower index constant at $n_1^2=1.0$, while the upper index is
changed. The results are shown in Fig. \ref{f1:3} and we
see that the first layer is well reconstructed for all choices of
$n_2^2$, while some fluctuations in the second layer are visible  when
$n_2^2>2$.

\begin{figure}[t]
  \centerline{\includegraphics[width=0.8\textwidth]{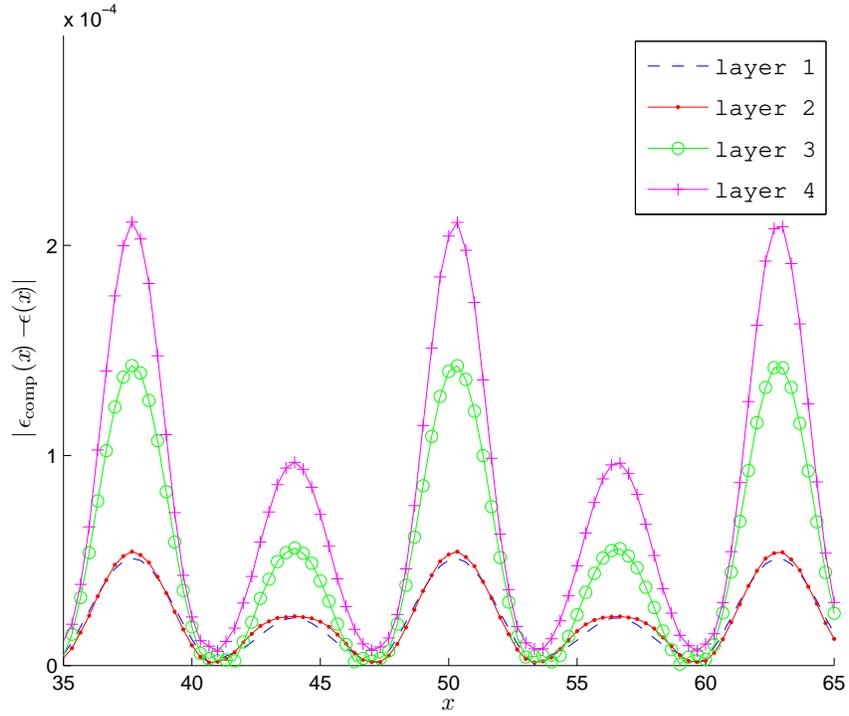}}
  \caption{The error in each layer, given as
  $|\epsilon_{\mathrm{comp}}-\epsilon|$. Parameters: $M=300$, $L=100$,
  $\omega \in [9,19]$,  
$N_{\omega} =100$, $\Delta z = \pi/2$, $n_1^2 = 1.0$, $n_2^2=1.05$. AKSFig5.eps.}
\label{f1:4}
\end{figure}

Finally, we demonstrate that it is possible to 
 reconstruct more than two
layers. Fig. \ref{f1:4} shows the error, given as
$|\epsilon_{\mathrm{comp}}-\epsilon|$ for each layer in a structure consisting of
four layers.  The
permittivity was chosen between $n_1^2=1.0$ and $n_2^2= 1.05$.

\subsection{Example 2, a square function}

In this example we have chosen  the permittivity to be  a  square
function with period $2 \pi$ in each layer. To get two different layers, and a permittivity between $n_1$
and $n_2$, we let  
\begin{eqnarray}
\epsilon_1(x) &=& 
\begin{cases}
n_2^2, & 0 \leq x < \pi \\
n_1^2, & \pi \leq x < 2 \pi,
\end{cases} \\
\epsilon_2(x) &=& 
\begin{cases}
n_1^2, & 0 \leq x < \pi \\
n_2^2, & \pi \leq x < 2 \pi.
\end{cases}
\end{eqnarray}

\begin{figure}[t]
  \centerline{\includegraphics[width=0.7\textwidth]{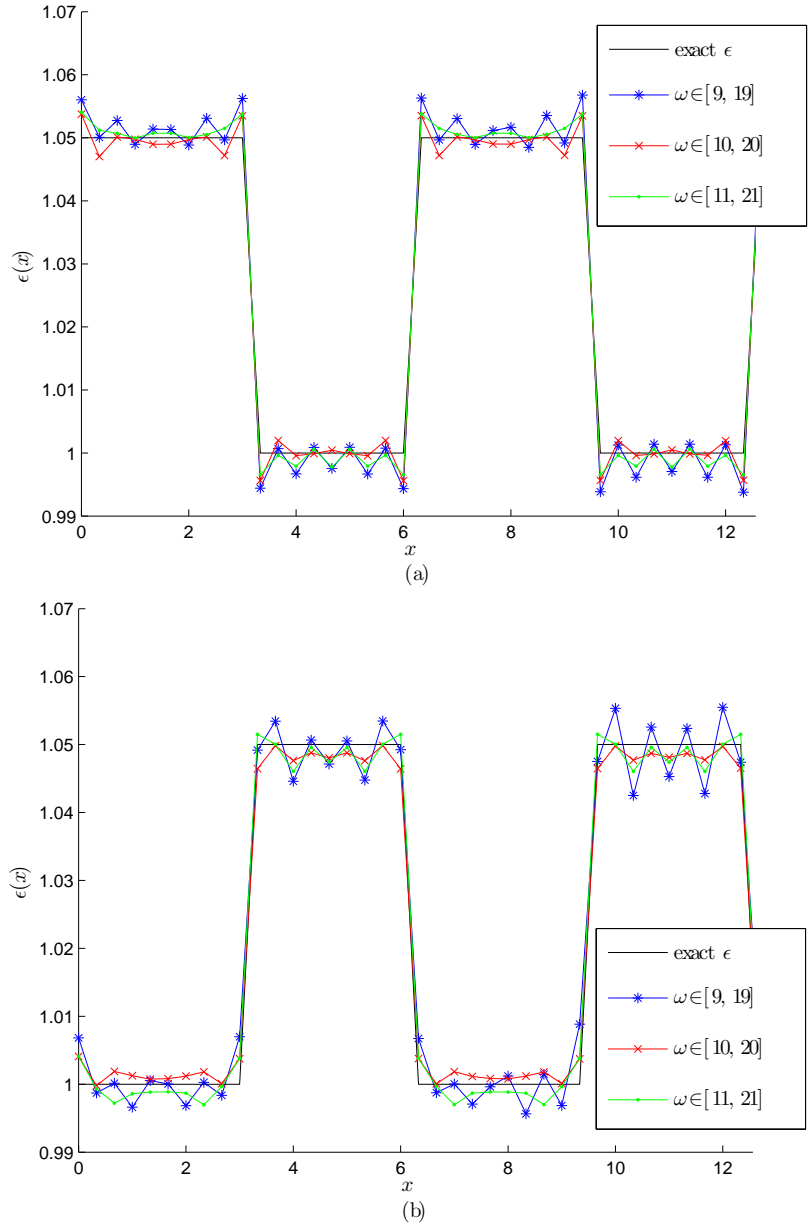}}
  \caption{Reconstruction of the square function for the first layer,
    (a), and for the second layer, (b). The computations were done for
    different $\omega_1$ 
  and $\omega_2$, but with constant bandwidth. Other parameters:
  $M=300$, $L=100$, $N_{\omega} =100$, $\Delta z = \pi/2$, $n_1^2 =
  1.0$, $n_2^2=1.05$. AKSFig6.eps.}
\label{f2:5}
\end{figure}

This example has been chosen to show that we are able to reconstruct a
less smooth function. The fact that the permittivity changes fast
seems to be in conflict with the discussion in Sec. \ref{num_ex}, but
as we see from Fig. \ref{f2:5} we actually manage to reconstruct two
layers of this structures as well. The reason it work is that we only
use the low frequency components of the square. 
The results can be seen in Fig. \ref{f2:5}.

\section{Conclusion}

In this paper we have presented a method for reconstructing 2d photonic
structures layer-by-layer. In principle, the reconstruction is exact,
and we have shown through numerical examples that we are able to
reconstruct different structures consisting of several layers. 

There are essentially three, fundamental mechanisms that limit the
accuracy in practice: The presence of evanescent modes, accumulated
reflection, and limited bandwidth. The fact that evanescent modes lead
to inaccuracies means that either the probing frequencies must be
sufficiently high, or the spatial transversal frequencies of the
structure must be sufficiently small. The second limitation is a result of the fact that
if the transmission through the structure is too small, little light
reaches the back end. Then the back end has little influence on the
reflection data, and cannot be reconstructed accurately. Finally,
the bandwidth must be sufficiently large such that the synthetic,
incident pulse is shorter than the round-trip time in all layers.
Alternatively, for a fixed, available bandwidth, the structure must vary
sufficiently slowly in the longitudinal direction $z$. In the numerical
examples, for a center wavelength \unit{1}{\micro\meter} the normalized frequency
interval $[9,19]$ corresponds to the wavelength interval \unit{[0.74, 1.56]}
{\micro\meter}. The layer thicknesses in the example become \unit{3.5}{\micro\meter}. If the
available bandwidth is reduced to $100$ nm, the layers must be 5--10
times thicker to achieve the same accuracy.

\clearpage

\clearpage

\clearpage

\clearpage

\clearpage

\end{document}